# The Properties and Origin
# of Kuiper Belt Object Arrokoth's Large Mounds

# Abbreviated Title:
# Kuiper Belt Object Arrokoth's Large Mounds

**S.A. Stern**
astern@swri.edu
Southwest Research Institute, Boulder, CO 80302

**O.L. White**
SETI Institute, Mountain View, CA, 94043
NASA Ames Research Center, Moffett Field, CA, 94035

**W.M. Grundy**
Lowell Observatory, Flagstaff, AZ 86001

**B.A. Keeney**
Southwest Research Institute, Boulder, CO 80302

**J.D. Hofgartner**
Southwest Research Institute, Boulder, CO 80302

**D. Nesvorný**
Southwest Research Institute, Boulder, CO 80302

**W.B. McKinnon**
Department of Earth and Planetary Sciences
and McDonnell Center for Space Science
Washington U. in St. Louis, St. Louis, MO 63130

**D.C. Richardson**
Dept. of Astronomy, U. Maryland, College Park, MD 20742

**J.C. Marohnic**
Dept. of Astronomy, U. Maryland, College Park, MD 20742

**A.J. Verbiscer**

**Dept. of Astronomy, U. Virginia, Charlottesville, VA 22904**

**S.D. Benecchi**
**Planetary Science Institute, Tucson, AZ 85719**

**P.M. Schenk**
**Lunar and Planetary Institute, Houston, TX 77058**

**J.M. Moore**
**Space Sciences Division, NASA Ames Research Center,**
**Moffett Field, CA, 94035**

**and the New Horizons Geology and Geophysics Investigation Team**

## Abstract

We report on a study of the mounds that dominate the appearance of Kuiper Belt Object (KBO) (486958) Arrokoth's larger lobe, named Wenu. We compare the geological context of these mounds, measure and intercompare their shapes, sizes/orientations, reflectance, and colors. We find the mounds are broadly self-similar in many respects and interpret them as the original building blocks of Arrokoth. It remains unclear why these building blocks are so similar in size— and this represents a new constraint and challenge for solar system formation models. We then discuss the implications of this interpretation.

## 1. Introduction

As a result of a flyby by NASA's New Horizons mission on 2019 January 01, the Cold Classical Kuiper Belt Object (CCKBO) (486958) Arrokoth became the first and remains the only KBO studied in detail by any spacecraft (e.g., Stern et al. 2019). It is a contact binary KBO, consisting of two bodies (referred to as "lobes") that apparently gravitated toward each other until they bonded (Stern et al. 2019; McKinnon et al. 2020); see Figure 1.

Amongst the most striking geologic features on the surface of Arrokoth are the ensemble of large, broadly similarly sized mound-like features seen on the larger lobe, Wenu. Though prominent, Arrokoth's large mounds (henceforth referred to below as "mounds" or "Arrokoth's mounds") have received scant attention in the literature, with only passing attention given in a handful of publications created to report broad results on this object (e.g., Stern et al. 2019; Spencer et al. 2020; McKinnon et al. 2020; Keane et al. 2022).

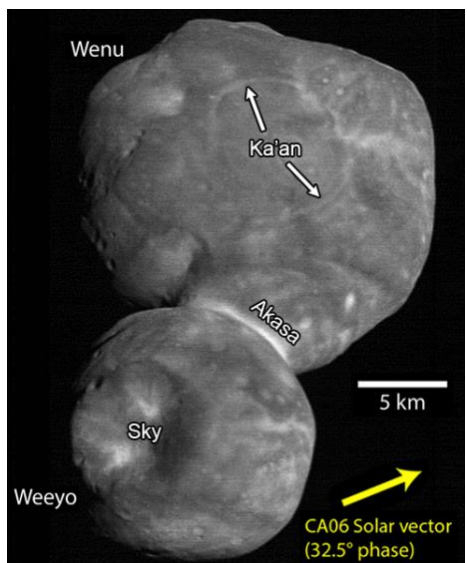

Figure 1. New Horizons CA06 LORRI image of Arrokoth depicting its construction from two lobes, Wenu and Weeyo. Wenu, the larger lobe, displays prominent mound units. Other named features that we refer to in the text are noted.

Here we intercompare all the available observational data on the individual mounds from New Horizons' LOng Range Reconnaissance Imager (LORRI; Cheng et al. 2008; Weaver et al. 2020) and its Multispectral Visible Imaging Camera (MVIC; Reuter et al. 2008; Howett et al. 2017). No other instruments on New Horizons resolved the mounds.

LORRI and MVIC observations provide the geological context, physical sizes and aspect ratios of the mounds, as well as their heights, areas, photometric, and color properties. LORRI and MVIC data also allow us to investigate evidence for mound-like structures on Arrokoth's smaller lobe, Weeyo. Below we also compare the mounds to large structural features seen by other spacecraft on short-period comets that formerly were KBOs before their accelerated evolution when they entered the warmer climes of the middle and inner solar system.

Finally, we offer and examine the hypothesis that mounds represent broadly self-similar, primordial building blocks of the KBOs and interpret this hypothesis in terms of its implications for Streaming Instability (SI) models of planetesimal formation in our outer solar system.

We begin with a discussion of the geological setting and context of the mounds.

## 2. Mapping and Geological Context

Two geologic maps of Arrokoth (Stern et al. 2019, Spencer et al. 2020) were produced previously. The first, very preliminary Stern et al. (2019) map was completed before downlink of CA06 LORRI, the highest-resolution (33 m/pixel, 32.5° phase) Arrokoth closest-approach imaging. It used the lower-resolution, CA06 MVIC observation as its base map (130 m/pixel, 32.5° phase). Stern et al. (2019) remarked on the possible assembly of Wenu by multiple discrete subcomponents (which they termed "rolling topography units"), although only eight mounds were identified at that time. By mid-2019, all closest-approach data had been downlinked, so Spencer et al. (2020) substituted CA06 LORRI as the base for an improved geologic map. The most significant modifications made to that map were on Wenu, which was divided into a "smooth plains" unit at the center of its face, surrounded by eight rolling-topography units. The improved resolution of this map allowed the contacts between the mound units, which in some cases had a vaguely interlocking configuration, to be traced with greater accuracy than before.

Although all useful data for geologic mapping of Arrokoth had been downlinked by the time the second map was made, we have produced an even-further-updated map for this study, since creation of the Spencer et al. (2020) map considered only the CA06 LORRI observation in isolation. We have since found that stereo imaging between CA06 LORRI and CA04 LORRI (138 m/pix, 12.9° phase), as well as the photoclinometric digital elevation models generated for the topographic investigation of Schenk et al. (2021), have provided a better understanding of Arrokoth's shape, and allowed us to define contacts now more confidently.

Here we produce a new version of the geologic map in two projections that use CA06 LORRI and CA04 LORRI as their base maps. There are some features within CA04 LORRI that are not apparent in CA06 LORRI (in particular, since it is at a lower phase, CA04 sees further around the limb of Arrokoth than CA06) and mapping CA04 therefore lends additional justification for how certain contacts in CA06 are mapped.

Figure 2 shows deconvolved versions of the CA06 and CA04 observations alongside the geologic maps overlain on each. Here we only map features that we can resolve on the relevant base map, so some features are mapped for the CA06 version that are not resolved in CA04, and so are omitted from its map. We now refer to the "rolling-topography units" as Arrokoth's large mounds, "mound units," or equivalently, mounds.

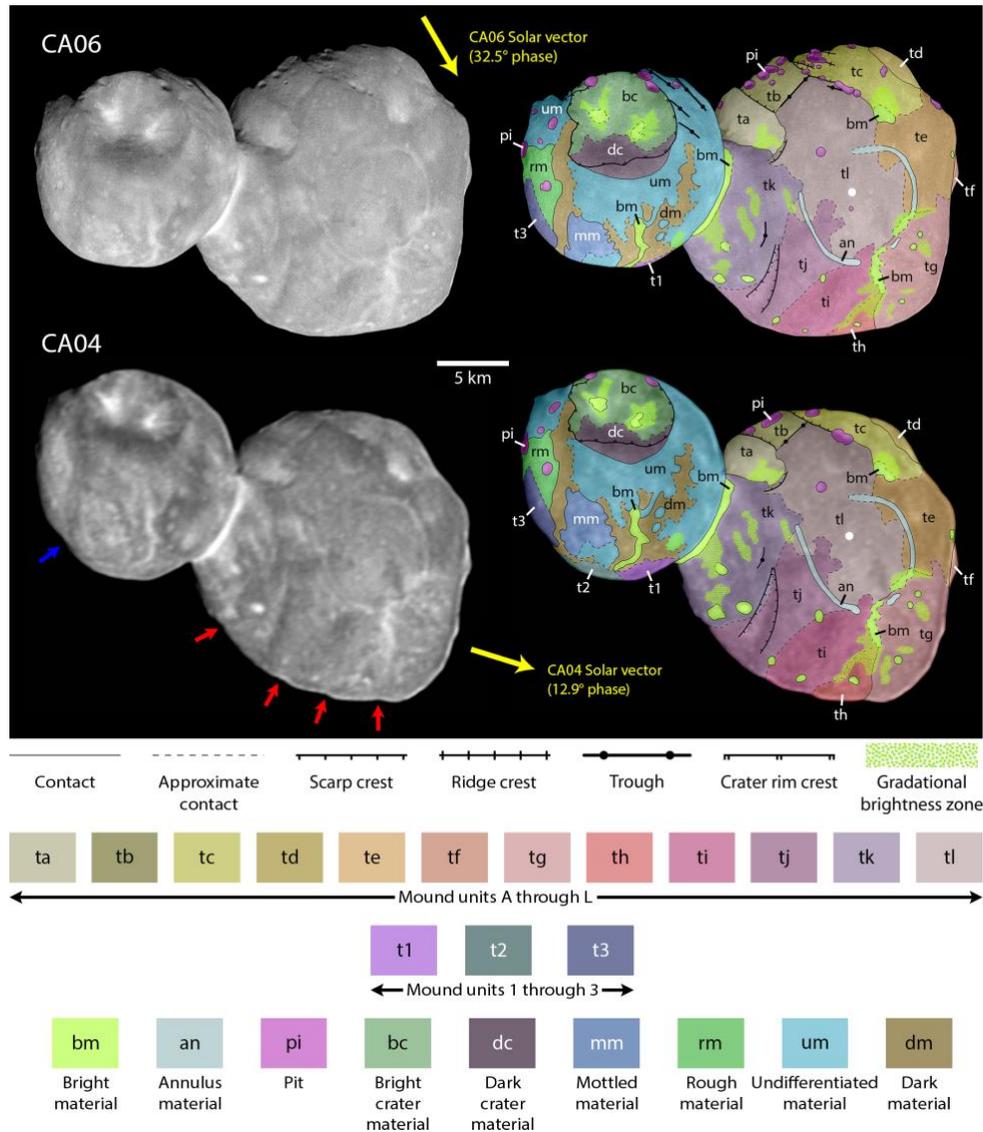

Figure 2. New Horizons CA06 LORRI and CA04 LORRI mosaics of Arrokoth (left) and the same observations overlain with geologic mapping (right). Mapped geologic units and lineations are defined in the legend. White dots on the geologic maps indicate the approximate location of the maximum principal axes of inertia for the larger lobe Wenu (Keane et al. 2022). If Wenu were once an isolated body, separate from Weeyo, this axis would correspond to its former spin axis. Blue and red arrows annotating the CA04 mosaic indicate notches on the limb of lobes Weeyo and Wenu, respectively.

Contacts between mounds are easier to identify close to the terminator owing to the illumination conditions accentuating topographic relief; their contacts often appear as troughs, scarps, and pit chains. Schenk et al. (2021) used photoclinometric digital elevation models to measure depths of a few 10s of meters for the troughs, and between 20 and 80 m for the pits. It follows that the mapping of contacts here has not been substantially modified from the map in Spencer et al. (2020), and units *ta* through *tg* are mostly unchanged as a result.

Contacts are harder to identify away from the terminator where the incidence angle is higher. Proximal to the limb, we have identified contacts where we observe a notch on the limb from which extends a faint lineation towards the center of Wenu's face. The CA04 observation has been crucial to this identification, as these notches are more apparent in CA04 than they are in CA06; red arrows in Figure 2 indicate their positions on the CA04 mosaic. These notches, which represent indentations on the limb of one pixel in the original CA04 observation, display relief of around 150 meters on the limb profile and are each about 1 km wide. They indicate that the generic boundaries between mounds manifest as shallow, V-shaped valleys; near the terminator, structural features such as the troughs, scarps, and pit chains sometimes punctuate these boundaries.

Additional evidence for the mound boundaries being locally depressed below adjacent terrain is that they are sometimes enhanced by a stretch of bright material (mapped as bright material, unit *bm*). Stern et al. (2019) and Spencer et al. (2020) each interpreted this material to be loose regolith that accumulated in local geopotential lows across Arrokoth. We maintain this interpretation. These accumulations provide a useful aid for mapping contacts and are especially bright where multiple boundaries converge and where adjacent slopes are especially high, such as at the neck connecting the two lobes of Arrokoth, and at the base of the topographically prominent mound *tg*.

As in the two previous iterations of the map, here we map the "gradational brightness zone" where the bright material has not accumulated to such a degree that it forms a consistently bright expanse but is instead spread thinly across the landscape. Where bright material does not mark contacts between mounds in the center of Wenu's face, we instead rely on albedo variations between the mounds, which may be caused by intrinsic albedo variation as well as subtle shading due to topographic relief. Consequently, contact definition on Wenu is most tentative within this central region.

We concur with Spencer et al. (2020) that the bright, incomplete, ~7.5 km diameter annulus near the center of Wenu's face, named Ka'an, is not a primordial boundary, owing to its smooth curvature, sharp definition, and the observation that it cuts across dark hills of units *te*, *ti*, and *tj*, all of which distinguish it from the more roughly defined and topography-conforming bright material. In fact, while it is plausible that the annulus is composed of the same bright material as unit *bm*, and even appears to intersect material of unit *bm* in one case, its unique configuration causes us to map it as a separate unit (annulus material, unit *am*) that crosscuts contacts between mounds and therefore overlies and is younger than them. It is plausible that this annulus of bright material is loose surface material that has moved downslope to this location (e.g., Keane et al. 2022).

Our revised geologic map identifies twelve mounds. Our new mapping of CA04 has caused us to remap the tentative boundaries subdividing the largest unit in the map of Spencer et al. (2020) as genuine contacts, and so we have split this unit into four mounds, including an additional one, *th*, that was not identified in Spencer et al. (2020).

The mounds appear to be organized such that smaller ones (units *ta* to *tk*) cluster around a large central mound (unit *tl*, previously mapped as "smooth plains" by Spencer et al. 2020). Wenu is estimated to be no more than ~10 km thick (Spencer et al. 2020), implying that these mounds, most of which are >5 km wide as seen in CA04 and CA06, may well extend all the way through the body of the lobe to the night side not seen by New Horizons. The fact that the inertial axis of the large lobe (Keane et al. 2022) is located within mound *tl* (white dots in the geologic maps) raises the possibility that if Wenu formed by low-speed accretion of these mounds within a spinning disk, then the smaller mounds were gravitationally drawn to, and accreted around, the equator of the largest central mound to assemble Wenu (see §7).

We derive a stratigraphic sequence of mound accretion on Wenu, conveyed in the "stratigraphic position" column of Table 1, which presents statistics for each of Wenu's twelve mounds. We assign numbers to each mound whereby a mound with a number higher than that of a mound bordering it accreted later. This sequence is derived based on three criteria, although only in a few cases do all three criteria apply to a single mound:

1. Crosscutting relations of mound contacts (whereby a mound accreting onto Wenu truncates an existing contact between two mounds that already accreted, creating a "T-junction", such as unit *ta* cutting across the contact separating units *tb* and *tl*);
2. The convexity of contacts (whereby mounds that accreted later form convex contacts with mounds that they superpose, such as those that units *tk* and *ti* form with unit *tj*); and
3. The topographic prominence of mounds relative to the center of Wenu as expressed by the angularity of their limb profiles (whereby mounds that accreted later have more of their original, bulk morphology exposed on Wenu's surface, such as the prominent and angular unit *tg*).

We gauge topographic prominence according to the convexity that a mound's limb profile presents in the CA06 observation, expressed as the angle subtended by the profile. mounds that are interpreted to be higher in the stratigraphic sequence generally show more convex profiles and are therefore more prominent.

Table 1 shows the number of pits mapped on each mound. We define a "pit" as an individual pit belonging to unit *pi* as seen closer to the terminator and as a bright spot (unit *bm*) as seen closer to the limb. These pits typically reach several hundred meters across. We concur with Spencer et al. (2020) and Schenk et al. (2021) that the bright spots represent accumulations of loose, bright material on the floors of pits that are equivalent to those seen near the terminator, but which display no topographic relief owing to the high incidence angle. Conversely, such bright material may also occur on the floors of the pits near the terminator but is invisible due to the low incidence angle.

Spencer et al. (2020) and Schenk et al. (2021) both mapped the locations of pits across Arrokoth, and our count is broadly similar to theirs, with most pits being identified near the terminator.

In total, we count 34 pits, some 7 of which are not assigned to a particular mound in Table 1, as they occur on contacts. The unequal distribution of pits across Wenu is likely due in part to the highly variable lighting conditions that make pits harder to identify nearer the limb, however, there is even substantial variability in pit distribution between those mounds along the terminator. Table 1 shows the areal fraction of each mound that is taken up by pits, and between units *ta* and *td* the fraction ranges from 0 to 0.177. The contrast between unit *ta*, with one pit and an areal pit fraction of 0.018, and neighboring pit *tb*, with seven pits and an areal pit fraction of 0.177, is especially striking. The cause of this unequal distribution is likely related to the origin of these pits, which may not be the same for all of them (Stern et al., 2019). As noted in Spencer et al. (2020) and Schenk et al. (2021), those pits that are in chains aligned with structural lineations (like the ridges within unit *tc*, and the scarp separating unit *tc* from *tl*) are likely endogenic in origin, resulting from surface collapse or venting along lines of weakness. Formation in this way densely concentrates these pits along the lineations.

Other pits that are scattered more sporadically are likely impact craters, and differences in pit spatial density between mounds may provide an indication of the impact environment during Wenu's accretion. The pits of units *ta* and *tb* are not obviously related to any structural lineations, and the very different pit counts of these mounds might be explained by them being differentially impacted prior to accretion, after which the heavily cratered unit *tb* abutted the less cratered unit *ta*. This would imply that the accretion process was sufficiently benign such that existing craters on the mounds were not obliterated. If the impact craters instead formed after accretion, the observed distribution would still imply steep gradients in impact rate across Arrokoth's surface, and the contrast between unit *ta*, which is proximal to and faces towards the neck, and other more pitted mounds along Wenu's terminator might be due to Weeyo shielding unit *ta* from impacts.

**Table 1. Basic Mound Properties on Wenu**

Statistics relating to each of the 12 mound units on Wenu mapped in this study. All measurements were obtained from the original CA06 LORRI observation and are uncorrected for projection. The three mounds that are viewed very obliquely in this projection, and which presented little surface area to the viewer, are highlighted in bold. Units *ta*, *tb*, *tc*, and *tl* do not show limb profiles, and so no limb profile convexity measurements were made for them.

| Mound | Area (km²) | Length (km) | Width (km) | Aspect Ratio | Convexity of Limb Profile (°) | Pits | Pit Areal Fraction | Stratigraphic Position |
|---|---|---|---|---|---|---|---|---|
| *ta* | 13.68±0.64 | 5.38±0.21 | 3.61±0.22 | 1.50±0.13 | - | 1 | 0.018 | 3 |
| *tb* | 9.32±0.59 | 4.05±0.21 | 3.03±0.18 | 1.34±0.12 | - | 7 | 0.177 | 2 |
| *tc* | 18.92±0.72 | 9.23±0.21 | 3.48±0.15 | 2.66±0.13 | - | 8 | 0.077 | 3 |
| **td** | **2.88±0.39** | **4.38±0.19** | **1.26±0.16** | **3.51±0.46** | **70.5** | **0** | **0** | **4** |
| *te* | 28.47±0.76 | 8.34±0.29 | 5.20±0.20 | 1.61±0.10 | 50.8 | 2 | 0.040 | 2 |
| **tf** | **0.64±0.37** | **3.00±0.20** | **0.63±0.21** | **5.10±1.75** | **38.4** | **0** | **0** | **3** |
| *tg* | 27.19±0.70 | 9.45±0.30 | 4.83±0.22 | 1.96±0.13 | 79.5 | 3 | 0.012 | 5 |
| **th** | **3.81±0.50** | **4.91±0.24** | **1.89±0.19** | **2.62±0.30** | **6.7** | **1** | **0.031** | **4** |
| *ti* | 22.02±0.91 | 8.97±0.19 | 3.42±0.17 | 2.62±0.14 | 19.8 | 1 | 0.006 | 3 |
| *tj* | 29.29±0.96 | 9.7±0.16 | 5.51±0.18 | 1.76±0.07 | 27.4 | 0 | 0 | 2 |
| *tk* | 45.25±0.79 | 9.82±0.21 | 6.45±0.21 | 1.52±0.06 | 27.7 | 2 | 0.029 | 4 |
| *tl* | 69.23±0.97 | 12.58±0.22 | 8.43±0.18 | 1.49±0.04 | - | 2 | 0.006 | 1 |

# 3. Possible Crypto-Mounds on Arrokoth's Small Lobe Weeyo

Wenu seems to be comprised entirely of mounds, raising the question of whether Arrokoth's smaller lobe, Weeyo, was constructed in a similar way. Stern et al. (2019) and Spencer et al. (2020) both described how Weeyo's geology is very different from Wenu's. This difference arises in part because Weeyo is not divided into subunits the way Wenu is in the CA06 observation. Instead, Weeyo displays complex albedo patterns across its surface and the effects of the large impact that formed the 6.7 km diameter crater, Sky. The latter has likely eliminated any accretional texture akin to Wenu's that Weeyo originally displayed, at least on its mapped face.

Yet by making use of stereo imaging in CA04 and CA06, we have tentatively identified three possible crypto-mounds on Weeyo's limb that are well-removed from Sky, and which may therefore be less affected by that impact/crater forming event. These mound units are labeled *t1* through *t3* in Figure 2, and they are most visible in the CA04 observation, which includes parts of Weeyo's surface further from Sky than in the CA06 observation. However, CA04's inferior resolution, and the fact that these mounds are only seen at an oblique angle on the limb, means that we cannot characterize these units as thoroughly as we can characterize the mounds on Wenu.

While these may be the best candidates for mounds on Weeyo, we acknowledge the possibility that they may instead be extensions of other mapped units, such as the undifferentiated material (unit *um*), dark material (unit *dm*), rough material (unit *rm*), and mottled material (unit *mm*). The least well-defined mounds are units *t1* and *t2*, which are only really visible in CA04 (a thin sliver of *t1* may extend into CA06) and are defined primarily by the observation that the limb of Weeyo bends at these two locations in a manner similar to how it bends at the apexes of prominent mounds on Wenu such as units *te*, *tf*, and *tg*. Foreshortening makes identification of mound contacts in the foreground of Weeyo difficult.

Unit *t3* is arguably the strongest candidate for a mound on Weeyo. This mound was originally mapped as part of unit *rm* in Stern et al. (2019) and Spencer et al. (2020), and unlike units *t1* and *t2*, much of it is visible in both CA04 and CA06. As with units *t1* and *t2*, it occurs at a bend in the limb profile of Weeyo, and a contact separating it from units *rm* and *dm* can be fairly confidently identified in both CA04 and CA06. In addition, a notch in the limb (indicated by the blue arrow on the CA04 mosaic in Figure 2) accentuates the contact of this mound with the dark material that borders it, in the same way that limb notches indicate mound contacts on Wenu. There is the possibility that the mottled material (unit *mm*), bounded mostly by dark material, may be a mound itself, but since no part of it falls along the limb, its morphology is obscure and this classification cannot be confirmed using available data. If these mound candidates on Weeyo are actually mounds, then their sizes (4 to 6 km wide as seen in CA04) are comparable to the smaller mounds on Wenu.

## 4. Basic Mound Properties

The basic sizes and shapes of Wenu mounds *ta–tl* and Weeyo crypto mounds *t1–t3* were characterized in the original CA04 and CA06 observations, uncorrected for projection. The polygonal area of each was determined from its vertices, and lengths and widths were estimated from the minimum-area bounding box that encapsulates all the mound's vertices. Consequently, the area measurements are more accurate than the length and width estimates. Uncertainties for all quantities are bootstrap estimates generated by varying the locations of the vertices by a few pixels and resampling.

Table 1 above lists the apparent area, length, width, and aspect ratio (i.e., the ratio of length to width) for the 12 mounds on Wenu as seen in the CA06 observation. Table 2 below presents the same information as seen in the CA04 observation alone, with the addition of measurements for the three crypto mounds on Weeyo. The only one of Weeyo's crypto mounds that is clearly visible in the CA06 observation is *t3*, which has an apparent area of 3.39±0.44 km$^2$, a length of 4.97±0.19 km, a width of 1.24±0.15 km, and an aspect ratio of 4.03±0.51. However, we note that these numbers are not exactly in agreement with those in Table 2 because values here are measured from the CA06 dataset, whereas the values in Table 2 are measured from the CA04 dataset. Crypto mound *t1* is viewed quite obliquely in CA06 with an apparent area of 0.41±0.23 km$^2$, a length of 2.89±0.20 km, a width of 0.72±0.14 km, and an aspect ratio of 4.15±0.90.

**Table 2. Additional Mound Properties on Both Wenu and Weeyo for the CA04 Observations**

| Mound | Area (km$^2$) | Length (km) | Width (km) | Aspect Ratio |
|---|---|---|---|---|
| Measurements of the 12 mound units mapped for Wenu and the three crypto mounds on Weeyo from the CA04 observation. Shown here are apparent sizes, uncorrected for projection. Five mounds that are viewed very obliquely in this projection and present little surface area to the viewer are highlighted in bold. | | | | |
| **Wenu** | | | | |
| *ta* | 9.39±1.81 | 4.19±0.28 | 2.89±0.27 | 1.45±0.17 |
| *tb* | 6.07±1.94 | 4.18±0.75 | 2.46±0.49 | 1.71±0.61 |
| *tc* | 15.73±2.56 | 9.60±0.33 | 2.84±0.23 | 3.38±0.31 |
| ***td*** | **1.99±1.63** | **4.45±0.30** | **0.86±0.24** | **5.19±1.49** |
| *te* | 24.63±2.50 | 8.72±0.40 | 4.72±0.29 | 1.85±0.15 |
| ***tf*** | **0.90±1.24** | **2.99±0.32** | **0.60±0.24** | **5.07±2.14** |
| *tg* | 32.60±2.90 | 9.77±0.41 | 5.10±0.28 | 1.91±0.16 |
| ***th*** | **8.46±1.92** | **5.66±0.31** | **2.84±0.26** | **1.99±0.21** |
| *ti* | 24.99±2.86 | 9.04±0.32 | 3.98±0.30 | 2.27±0.19 |
| *tj* | 34.92±3.38 | 9.96±0.27 | 6.32±0.28 | 1.58±0.08 |
| *tk* | 49.59±3.39 | 10.37±0.35 | 6.54±0.32 | 1.59±0.10 |
| *tl* | 63.31±3.62 | 12.86±0.40 | 7.33±0.29 | 1.76±0.10 |
| **Weeyo** | | | | |
| ***t1*** | **3.02±1.41** | **4.46±0.31** | **1.29±0.25** | **3.48±0.73** |
| ***t2*** | **2.41±1.41** | **4.22±0.31** | **1.02±0.25** | **4.15±1.07** |
| *t3* | 4.37±1.85 | 5.25±0.31 | 1.32±0.23 | 3.99±0.75 |

Figure 3 compares the apparent areas, aspect ratios, lengths, and widths for each of the mounds/crypto-mound candidates measured from CA04 and CA06. Uncertainties are lower for the CA06 measurements because the spatial resolution of CA06 is ~4 times better than that of CA04. The areas and, to a lesser extent, the lengths and widths of mounds that are well-observed in both datasets are consistent within the reported uncertainties, and the apparent aspect ratios of the mounds on Arrokoth are all ~2. Some disagreement between the two observations is expected since the viewing geometry changes between them (Figure 2) causing some mounds/crypto-mounds to be viewed more clearly in CA04 than CA06, and vice versa. Generally speaking, mounds *t1–t3* and *tg–tk* are better viewed in CA04, and mounds *ta–tf* and *tl* are better viewed in CA06.

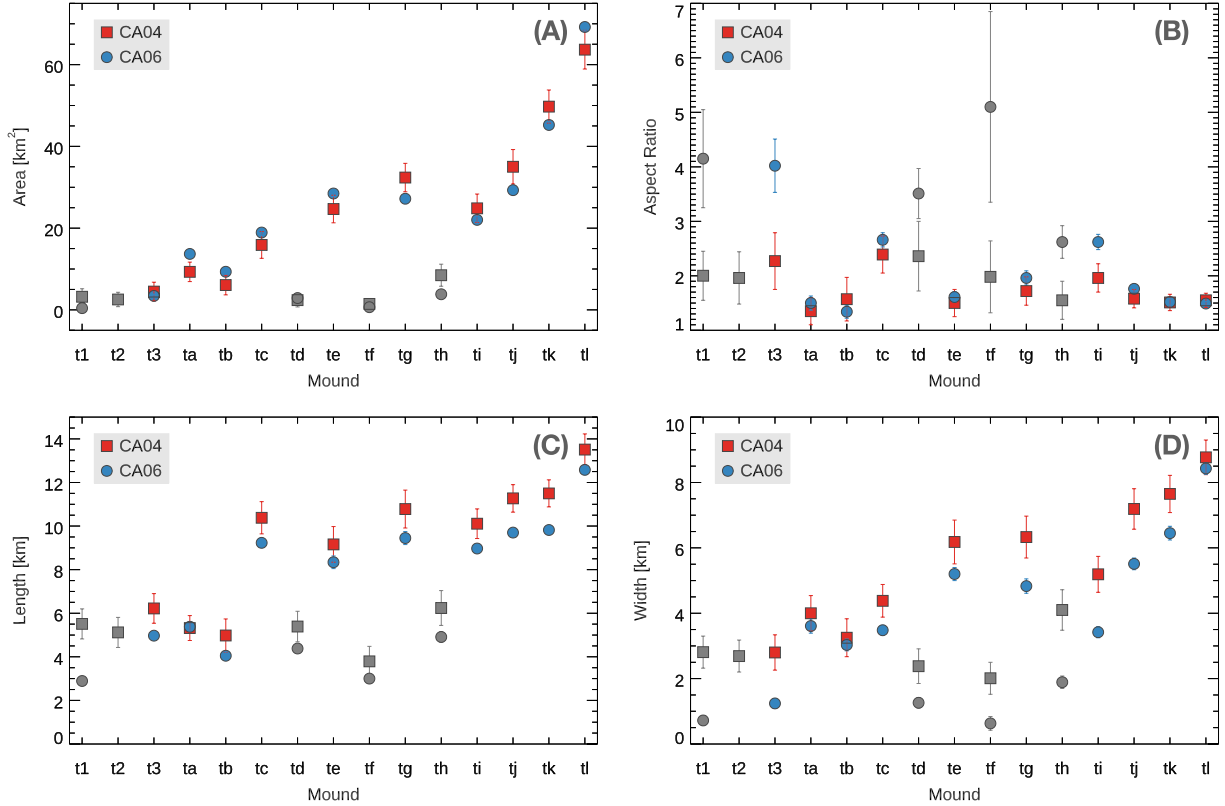

Figure 3. Comparison of mound properties measured from New Horizons CA04 (squares) and CA06 (circles) observations of Arrokoth. The apparent area, aspect ratio, length, and width are shown in Panels A–D, respectively. Gray points denote mounds that are viewed obliquely and present little surface area to the viewer.

## 5. Mound Photometry and Photometric Comparisons

We use the Hofgartner et al. (2021) normal reflectance map of Arrokoth to investigate the photometry of Wenu's mounds. Normal reflectance is equal to $I/F$ when both the incidence and emission angles are zero, where $I$ represents the scattered intensity and $\pi F$ is the solar flux at the solar distance of the scattering surface. Normal reflectance therefore measures the intrinsic brightness of a surface. The Hofgartner et al. (2021) map was produced by fitting a lunar-Lambert photometric function to New Horizons panchromatic LORRI images and previously measured Hubble Space Telescope absolute magnitudes (Benecchi et al. 2019a; 2019b) of Arrokoth. The normal reflectance map is at the LORRI pivot wavelength of 607.6 nm (see Weaver et al. 2020).

Pixels within each mound were extracted from the normal reflectance map using the boundaries shown in Figure 2. Pixels mapped as bright material, annulus material, pit, and gradational brightness zone were excluded. Pixels corresponding to incidence or emission angles >89º (terminator or limb) were also excluded because the uncertainties of the normal reflectance map are greatest where these angles approach 90º. The normal reflectance means and the standard deviations for each mound are given in Table 3. Mounds *td*, *tf*, and *th*, have been excluded because of their smaller observed areas (Figure 3).

Figure 4 shows the normal reflectance cumulative probability for each mound. Overall, all of Wenu's mounds are photometrically similar, with similar normal reflectance means and distributions. The mounds are also photometrically similar to Arrokoth's global-average photometry. Differences between mounds are small and we discuss the most prominent of the subtle differences identified. Mounds *ta–tc* have greater normal reflectance standard deviations than other mounds, which results from their location near the terminator where normal reflectance errors are largest. Indeed, we verified that as the imposed pixel cut based on incidence angle becomes stricter (decreases from 89º), the standard deviations of mounds *ta–tc* become more similar to those of the other mounds. mound *tg* has a normal reflectance mode slightly greater than the other mounds, possibly related to its suggested later emplacement (see §2 above). Finally, mound *tb* has the least Gaussian normal reflectance distribution, due to an excess of modestly darker-than-average pixels. This difference is attributed to a region of mound *tb* near its boundary with mounds *ta* and *tl*, that has intrinsically lower normal reflectance than the average on Wenu. Mound *tb* has fewer pixels than most other mounds so the effect of this region is greater for the normal reflectance distribution than it would be on other mounds. Thus, this difference may not be indicative of a different history for mound *tb*.

**Table 3. Photometric Statistics of Wenu's Mounds.**

| Mound | Normal Reflectance Mean | Normal Reflectance Standard Deviation |
|---|---|---|
| *ta* | 0.259 | 0.096 |
| *tb* | 0.247 | 0.097 |
| *tc* | 0.238 | 0.072 |
| *te* | 0.243 | 0.033 |
| *tg* | 0.263 | 0.022 |
| *ti* | 0.246 | 0.023 |
| *tj* | 0.242 | 0.030 |
| *tk* | 0.237 | 0.034 |
| *tl* | 0.243 | 0.028 |

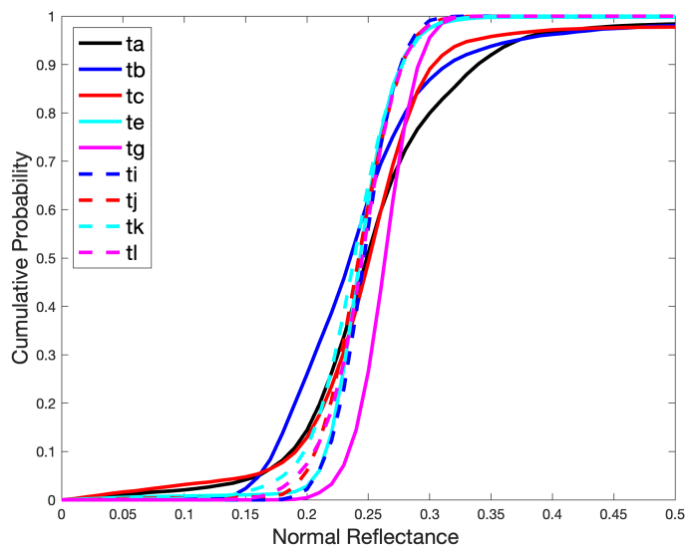

Figure 4. Cumulative normal reflectance distributions of Wenu's nine largest (i.e., most aerially extensive) mounds. Overall, all are photometrically similar. Differences are discussed in the text.

# 6. Mound Colors and Color Comparisons

New Horizons' MVIC color imager has three broadband color filters: "BLUE," covering 400 to 550 nm, "RED," covering 540 to 700 nm, and "NIR," covering 780 to 975 nm. MVIC also has a narrow "CH$_4$" filter covering a methane-ice absorption band from 860 to 910 nm, but, owing to its narrow bandpass, this filter provided relatively low signal to noise at Arrokoth and will not be further discussed here.

Each filter is affixed to a CCD array with 5000×32 active pixels. The four CCDs are arranged adjacent to each other in MVIC's focal plane. They are operated in time delay integration (TDI) mode, with the set of four arrays being scanned across the target scene at a rate corresponding to the readout rate across the short axis of each array. This operation mode results in the accumulation of a four-color composite image that is 5000 pixels wide with a length set by the duration of the scan and the scan rate.

New Horizons' highest-resolution MVIC color observation of Arrokoth was CA05, obtained at 05:15 UTC from a mean range of 17,200 km, between the times of the panchromatic (LORRI) CA04 and CA06 observations described earlier. The image scale was 340 m/pixel, though the actual spatial resolution was coarser than that, owing to the instrumental point spread function as well as changing spacecraft-target geometry during the observation. To avoid resampling the data, we co-registered the separate wavelengths to Arrokoth using nearest-integer shifts, based on centroids computed in the two axes. The registered image is shown in Figure 5a.

Specific regions of interest (ROIs) were selected to exclusively sample each of the larger mounds on Wenu, avoiding pixels near the limb or near mound boundaries where image smear could lead to pixels being contaminated by space or by another mound. We also avoided areas mapped as stippled in the geological map, since these appear to have been modified by more recent processes that could have altered their colors. These ROIs are shown in Figure 5b. The number of uncontaminated MVIC pixels selected varied by mound, ranging from as few as 10 for *ta* to over 200 for *tl*.

A comparison of color ratios among the ROIs shows that most of the mounds are similar in terms of average color and color distribution. The mounds are also generally similar in color to Arrokoth's global average. For mounds *tg*, *ti*, *tj*, *tk*, and *tl*, mean NIR/RED I/F color ratios range from 1.50 to 1.52, all with variances of 0.04, as shown in a cumulative plot (Figure 6). RED/BLUE colors are noisier, but similarly distributed with I/F color ratios of these mounds ranging from 1.43 to 1.50 with variance from 0.08 to 0.11. BLUE, RED, and NIR colors can be combined into a color slope. Following the convention used for KBO visual wavelength colors in the literature, this slope is called *S* and is given in units of percent rise per 100 nm relative to the anchor wavelength of 550 nm. For mounds *tg*, *ti*, *tj*, *tk*, and *tl*, mean color slopes range from 26.6% to 27.6% per 100 nm, with variances from 2 to 3 in the same units. These colors are within the normal range for CCKBOs (e.g., Hainaut & Delsanti 2002; Pike et al. 2017; Grundy et al. 2020).

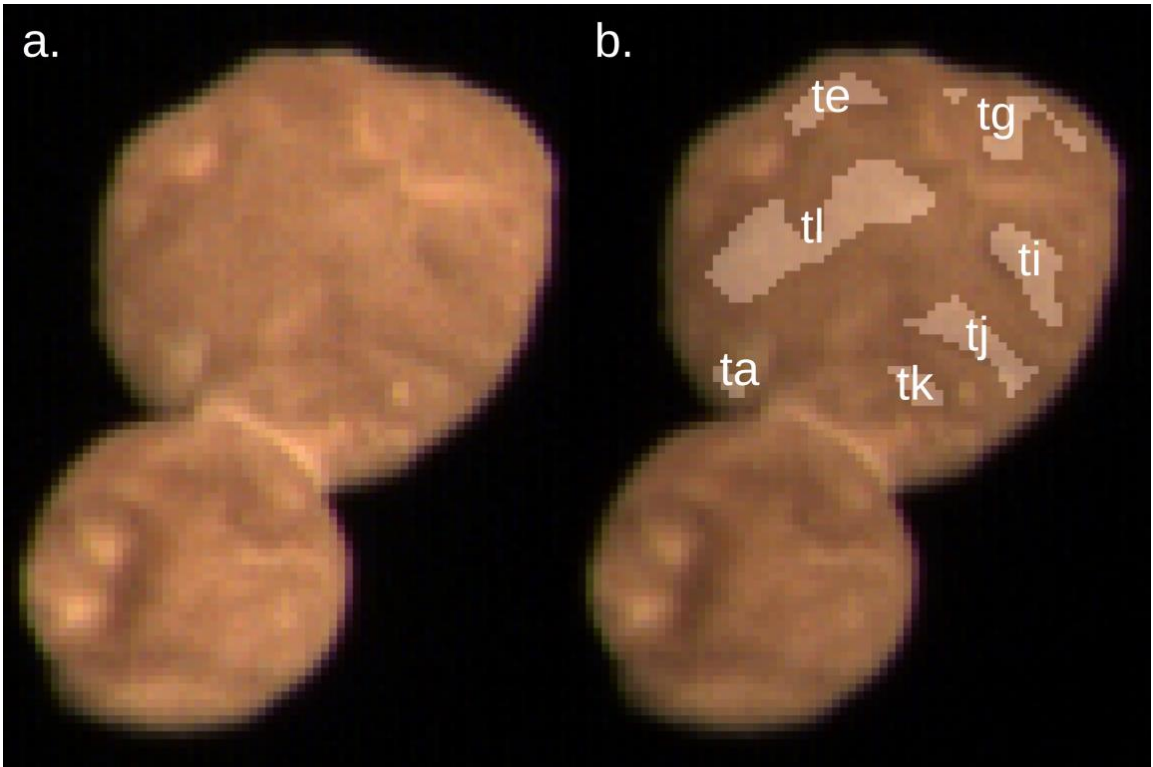

Figure 5. (a) CA05 MVIC color observation of Arrokoth, with three MVIC broadband filters registered to nearest-integer pixel and displayed in order of increasing wavelength with 400–550, 540–700, and 780–975 nm filters displayed in blue, green, and red, respectively. (b) Same, with pixels selected for inclusion in individual mound ROIs highlighted and labeled.

Two of the mound ROIs stand out as having colors distinct from the others. The largest difference is seen on *ta*, which is less red than the others at near-infrared wavelengths, but redder at BLUE wavelengths (NIR/RED ratio 1.39, RED/BLUE ratio 1.53, *S* slope 24.4). However, *ta*'s proximity to the neck region suggests its distinctive color could be associated with whatever process is responsible for the bright neck, rather than that mound being composed of distinct material. Mound *te* is slightly redder (NIR/RED ratio 1.53, RED/BLUE ratio 1.49, *S* slope 28.3). These color differences appear to be intrinsic to *te* rather than the result of subsequent modification.

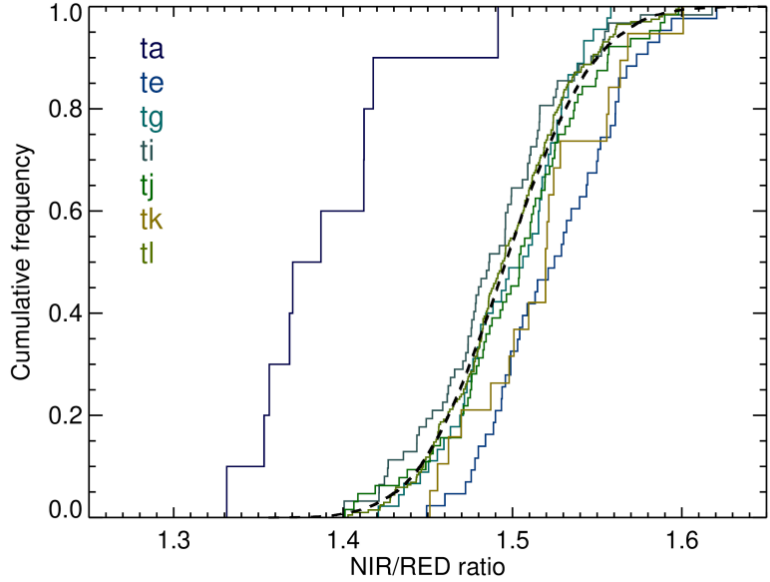

Figure 6. Cumulative color distributions for the mound ROIs. A redder color distribution plots further to the right. The dashed black curve simulates the effect of MVIC's known noise characteristics for mound *tl*, indicating that the apparent breadth of the mound's color distribution is entirely attributable to image noise.

The variance in color ratio for an ROI might give an indication of the distribution of colors within that unit, but image noise will also contribute to an apparent distribution of colors even in a region with perfectly uniform colors. To test for this possibility, we used MVIC's known noise characteristics to generate random colors around the mean color of mound *tl*, the one with the most available pixels. The resulting color distribution is shown with a dashed curve. It almost exactly matches the observed color distribution of *tl*, indicating that all the observed color variation within that ROI can be attributed to noise and we should assign no physical significance to the variances of the observed color distributions.

The significance of the colors can be assessed via a two-tailed Kolmogorov-Smirnov comparison between the color distributions. NIR/RED colors of *ta* are distinct from the colors of *tl* at 5.3σ confidence, and colors of *te* are distinct from those of *tl* at 3.1σ confidence. Colors of the other mounds are all consistent with having been drawn from the same parent distribution.

## 7. Mound Properties Synthesis and Comparative Planetology

The discussion above reveals a remarkable self-similarity in the sizes and axial ratios of the well-observed mound structures that comprise Wenu. Their photometric and color properties are generally self-similar as well. Such broad correspondence between the observable properties of all the Wenu mounds suggests a common origin and evolution.

These kinds of large mound structures have not, to our knowledge, been exhibited on other small bodies. The two lobes of comet 67P/CG at 4.1 km×3.3 km×1.8 km and 2.6 km×2.3 km×1.8 km are, however, somewhat similar in length scale to the smaller mounds of Arrokoth. For example, the largest two dimensions of the larger lobe of comet 67P/CG (4.1 km×3.3 km) are not very different from the observed length and width of mounds *tb* (4.0 km×3.0 km) and *tf* (3.0 km×0.6 km). Larger comets might be accumulations of smaller mound units or they could be mound units similar in scale to Arrokoth's larger mounds or both.

Layers have, however, been observed on at least two comets (Thomas et al. 2007; Massironi et al. 2015), and the *talps* model (*splat* spelled backwards) for these layers predicts an increase of layer size with distance from the center of the nucleus (Belton et al. 2007). An intriguing possibility is that Wenu is similarly layered and its mounds correspond to layers much larger than those observed on comets because they are much further from Wenu's center.

It is therefore possible that observed small comet nuclei, whether single lobed (e.g., 81P/Wild 2 or 9P/Tempel 1) or bilobate (e.g., 1P/Halley, 19P/Borrelly, 67P/Churyumov-Gerasimenko), are single- or double-mound structures. Conversely, however, the lack of apparent mound subunits on the surfaces of individual cometary nuclei explored by spacecraft may be a result of differences in formation location and evolutionary history, particularly as all comets observed by spacecraft have undergone rapid, thermally driven, morphological evolution due to entering the inner solar system numerous times. Furthermore, some models predict that most observed comets are collisional fragments or gravitational aggregates formed in collisional events (Bottke et al. 2023), processes minimized among the cold classical KBOs; we therefore discount this latter possibility for Arrokoth. We do, however, recognize that searches for and the study of features that may be formation subunits on cometary nuclei imaged by spacecraft would be of interest, though this is beyond the scope of this Arrokoth paper.

# 8. On an Exogenic Mechanism for Mound Formation: Initial Accretion Simulations

Previous accretion studies addressed the bilobate shape of Arrokoth and found that the merger between the large and small lobes must have been slow (<5 m/s) and grazing (impact angle >75°) to leave intact lobes after the merger (e.g., Stern et al. 2019, McKinnon et al. 2020, Marohnic et al. 2021). This is consistent with the two lobes having a close synchronous orbit before the merger (McKinnon et al. 2020). After the merger, the rotation period of Arrokoth should have been equal to the mutual orbital period of the two lobes immediately before the merger. The current rotation period of Arrokoth is very close to 15.9 hours (Spencer et al. 2020). This would match the orbital period if the two lobes have very low bulk density (~0.25–0.29 g/cm³; Spencer et al. 2020, McKinnon et al. 2020)[1]. Here we address the oblate shape of the large lobe of Arrokoth (Wenu) and its mounds.

---

[1]Note that Spencer et al. (2020) used spheres to calculate the density while McKinnon et al. (2020) used ellipsoids.

We performed simulations with the soft-sphere version of the PKDGRAV code (Richardson et al. 2000, Stadel 2001, Schwartz 2013 Zhang et al. 2017, 2018). PKDGRAV is a parallelized *N*-body gravity tree code that models persistent frictional contacts between spherical particles. In this code, the equations of motion are integrated using a second-order leapfrog algorithm with a fixed timestep chosen so that interparticle contacts (represented as small overlaps between the particles as a proxy for surface deformation) do not exceed 1% of the smallest particle radius. Collisions are resolved using a spring-dashpot model parameterized by Hooke's Law spring constants, a shape value, a set of static, sliding, rolling, and twisting friction constants, and (optionally) interparticle cohesion. The cohesion and other parameters of PKDGRAV particles were set to reproduce the behavior of gravel (e.g., the angle of repose ~34 deg; Zhang et al. 2018).

The simulations we undertook tested two different formation models. In the first model, we tested whether the mounds can be produced by low-speed impacts ("splashes") of ~3 km diameter bodies on the large lobe. Wenu would have to form by some unspecified mechanism before the simulated splashes. We found that splashes can produce mound-like mounds only for very low impact speeds (<1 m/s). For impact speeds above ~1 m/s, the projectile material is excessively flattened and spread over too large an area of the target surface to resemble the mounds.

The surface pattern obtained in successive <1 m/s splashes of ~3 km projectiles is complex. The mounds from the last few impacts are left fully exposed. The mounds that formed previously are partially or almost fully buried by subsequent splashes. This creates complex topological terrains where the exposed areas of mounds vary between 0% (fully buried) to 100% (fully exposed), and everything in between. The exposed mound areas form a fragmented (stochastic) pattern. For comparison, the mounds on the large lobe of Arrokoth seem to be organized in a more orderly fashion and are more similar in size. We leave a detailed study of this model for future work.

In the second model, we tested whether the mounds could be what remained of individual bodies from which the large lobe accreted. First, a progenitor object 5 km in diameter (not likely spherical, more likely flattened) was created by producing a randomly distributed cloud of 1559 PKDGRAV particles with no relative spin or velocity. The cloud was then allowed to collapse under its own gravity in a separate, preparatory simulation and allowed to settle to a steady state. Due to the collapse mechanism, the particles in the body were randomly packed rather than arranged in an orderly manner that could confer additional, unrealistic (non-physical) strength. Thirty copies of the body resulting from this collapse simulation were then used as the progenitor objects in the Wenu formation model simulations described in greater detail below. The cohesion and other parameters of PKDGRAV particles were set to reproduce the behavior of gravel (e.g., angle of repose ~34 deg; Zhang et al. 2018). The bulk density of all bodies was set to 0.25 g/cm$^3$.

The 30 bodies were initially randomly distributed in an ellipsoidal volume with 35 km long axes and 17.5 km short axis (i.e., an oblate ellipsoid with 2:1 flattening). The minimum separation between surfaces of two neighbor bodies was 2.5 km. The whole cloud was given slow rotation around the short axis (corresponding to a ~100-hour period). The simulation was run with an adequately small timestep for two days of simulation time (about one day of real time on 28 Broadwell cores of NASA's Pleiades supercomputer).

Figure 7 shows three snapshots from this simulation and Figure 8 compares the final result from this simulation of Wenu. The adopted global rotation is not fast enough to prevent gravitational collapse and accretion of progenitor bodies into a single object. The volume of the final object is by design similar to that of Wenu (Spencer et al. 2020). The final object is an oblate axi-symmetrical ellipsoid (22.2 km × 21.2 km × 10.8 km; with roughly 2:1 flattening) with a ~0.2 g/cm$^3$ bulk density and 20.4-hour rotation period. This favorably compares with the overall shape and rotation period of Wenu (21.2 km × 19.90 km × 9.05 km; Keane et al. 2022).

The resulting final shape is a direct consequence of the assumed global rotation of the cloud. If the initial rotation period is >3 times longer than the one assumed in the nominal simulation, the flattening of the final object is small (<1.5 axial ratio). If the initial rotation period is >3 times shorter, the flattening of the final object is excessive (>3 axial ratio). Wenu's shape therefore informs us about the properties of the original group of bodies from which it formed (in the context of this model). We can also rule out radial and/or random velocities above ~0.5 m/s; the accreted object ends up being significantly less flattened than Wenu in these cases. Overall, these results suggest a final stage of Wenu's accretion arising from a rotating group of ~5 km bodies that gently settled together.

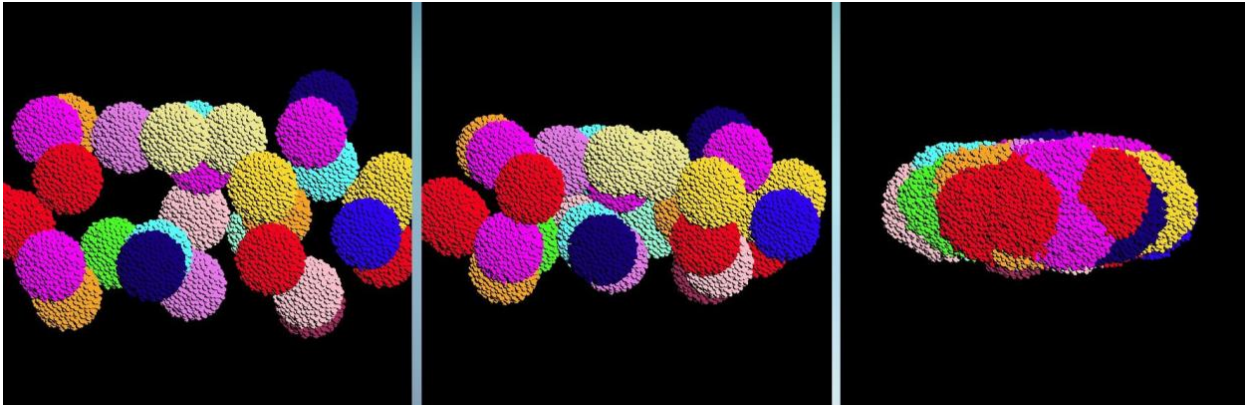

Figure 7. Three snapshots from the nominal PKDGRAV simulation of Wenu's accretion from 30 progenitors ~5 km in diameter.

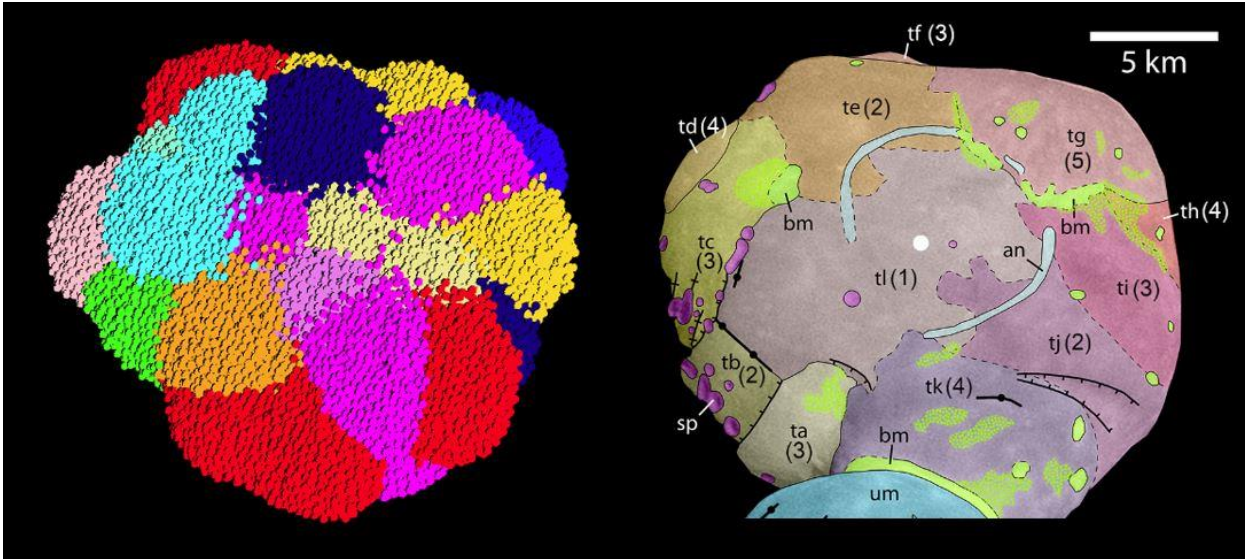

Figure 8. Comparison of the final object obtained in the nominal simulation with the CA06 version of the geological map of Wenu from Figure 2. The smaller central mounds in the simulation contrast to Arrokoth itself, which displays its largest mounds at the lobe center. This testifies to the need for future work.

The object produced in our nominal simulation shows mounds with the scale and distribution that favorably compare with observations (Figure 8). The mounds are exposed areas of the original bodies that maintained their integrity and were not quite flattened during their gentle accretion. As all bodies accrete, they compete for space, and this leads to a situation where the exposed areas are somewhat similar in size (we assume here that the progenitor bodies were similar in size), and they are arranged in an orderly fashion on the surface. As with the observed mounds, the simulated mounds display an identifiable stratigraphic sequence based on crosscutting relations of the mounds, the convexity of mound contacts, and mound topographic prominence as described in section 2. Some of the simulated mounds may be too prominent compared to the more leveled surface of Wenu, but this problem could be resolved if the surface of Wenu was slightly disturbed (e.g., by subsequent small impacts). Modest radial velocities of the progenitor bodies would also reduce the prominence of individual mounds.

The main assumption of the model described here is the formation of ~5 km progenitors in a rotating, gravitationally bound cloud that is subject to gentle collapse. Here we can only speculate on how this fits with more general ideas about planetesimal formation. Arrokoth is a part of distant dynamical CCKBOs which are thought to have formed in situ at 40–47 AU, where CCKBOs remained largely dynamically and collisionally undisturbed. They thus represent the best constraint that we have on planetesimal formation (Nesvorný et al. 2022).

Stern et al. (2019) and McKinnon et al. (2020), among others, argue that CCKBOs formed by the Streaming Instability (SI, Youdin & Goodman 2005, Nesvorný et al. 2019), an efficient mechanism to concentrate ~cm size particles (pebbles) in a gas nebula. High-resolution SI simulations show the formation of large pebble clouds that undergo gravitational collapse. Previous studies indicated that the collapsing pebble clouds produce planetesimals with the characteristic size of CCKBOs (Simon et al. 2017, Kavelaars et al. 2021). Planetesimals are often found to form in binaries in this model with characteristics that match the binaries observed among CCKBOs (Nesvorný et al. 2010, 2019; Grundy et al. 2019). Arrokoth itself—a finely balanced contact binary KBO—provides direct evidence for planetesimal formation by the SI (McKinnon et al. 2020; as opposed, for example, to random accretion of bodies by energetic collisions).

Previous studies of gravitational collapse were not able to resolve objects as small as mounds (~ 5 km). The model's spatial resolution is a fundamental problem of SI collapse simulations. For reasons related to resolution, we are therefore unable to directly link previous SI collapse models to the soft-sphere PKDGRAV simulations described in this work. Instead, we attempt to reverse-engineer the conditions that may have existed during the last (so-far-unresolved) stages of collapse. The best results are obtained when ~5 km progenitors are distributed in a rotating, gravitationally bound cloud that is subject to a gentle collapse. Our expectation is that future SI studies with improved spatial resolution will be able to validate these results (or rule them out, for example, if the mound-size objects do not show up). Below we briefly discuss several considerations related to this scientific problem.

For simple collapse, where the whole cloud of pebbles collapses as a single unit, planetesimals should form directly from pebbles; hence they would presumably have smooth surfaces and no mounds. There is a possibility, however, that the initial perturbations of density and velocity fields of pebbles, inherited from the turbulent gas nebula, were amplified during the collapse, and first led to the accretion of intermediate-scale bodies. Moreover, the collapsing cloud, where inelastic collisions between pebbles efficiently damp random velocities, could be subject to gravitational fragmentation on various spatial scales.

Finally, we point out that a pebble cloud, if it stays near virial equilibrium, would not be subject to gravitational fragmentation. Real clouds, however, are expected to slowly free-fall (Nesvorný et al. 2020), too slowly for virial equilibrium to be established. This should lead to a situation where the Jeans mass becomes progressively smaller and the pebble cloud undergoes complex gravitational fragmentation. In this context, the hypothetical ~5 km progenitors of Wenu could arise from the shortest unstable wavelengths. Whereas the existing numerical simulations do not have a sufficient resolution to test these ideas in detail, they show the fundamental role of initial rotation and its cascading to progressively smaller spatial scales (e.g., Robinson et al. 2020, Nesvorný et al. 2021).

## 9. Mound Origin Discussion

We find that the individual mapped mounds on Arrokoth's larger lobe, Wenu, are consistent with the merger or assembly of discrete, similarly sized multi-km-scale planetesimals from Arrokoth's natal collapse cloud. As described just above, our numerical calculations of collisional mergers of precursor bodies indicate that normal impact speeds ≲1 m/s are necessary to preserve the shapes of the individual subunits, using gravel friction parameters.

Low cohesion of the individual mound precursor units is apparently mechanically necessary at the time of the Wenu merger collisions as well. Otherwise, the shape of Wenu would much more reflect the shapes of the individual subunits (mounds) from which it was built (i.e., it would resemble a bunch of grapes). This is not to say that there is no structural trace of the mounds. Wenu itself is an oblate body roughly hexagonal in equatorial outline (Keane et al. 2022, Porter et al. 2023) and this shape's "lumpiness" corresponds directly to the mounds/subunits we have been discussing.

Wenu's overall flattening (0.60; Porter et al. 2023) provides a possible indicator of its original rotation rate, prior to its later merger with its junior partner, Weeyo, Arrokoth's smaller lobe. For a uniform density of 0.25 (0.50) g/cm$^3$, and for the assumption that Wenu overall represents an equilibrium figure, this flattening implies an initial, post-mound-assembly rotation period of 12.7 (8.9) hrs. Alternatively, as we have discussed above, Wenu's shape may instead be a direct inheritance from the aspect ratio and rotation rate of the cloud of mound-forming subunits prior to final accumulation.

But why did a set of broadly similar ~5-km diameter mound-forming bodies assemble to form Wenu, as opposed to forming Wenu from a hierarchy of building-block sizes, or simply from a vast assembly of small pebbles? The latter is the usual conceptual picture for planetesimal formation via the Streaming Instability (SI) or other aerodynamic concentration mechanism. The highest-resolution pebble cloud evolution study to date (Nesvorný et al. 2021) shows clouds collapsing to dense, rotating disks, followed by spin up and mass loss to spiral arms. Determining whether a characteristic planetesimal building-block size emerges from further centrifugal-gravitational fragmentation in such situations is an open question requiring more simulations at even greater resolution.

And how generalizable would any such planetesimal building block size be? Presumably, Weeyo, while smaller and less oblate, was assembled from similar subunits as Wenu, but evidence for mound structure on Weeyo is less certain and harder to discern, as we have discussed, possibly because of the obscuring effects of the dominating crater Sky's formation.

Streaming Instability simulations generally result in a population of different-sized planetesimals, albeit sometimes with a characteristic size or mass (e.g., Johansen et al. 2015, Simon et al. 2016). For the cold classical belt, the characteristic size is ~100-km diameter, based on telescopic data (Fraser et al. 2014), and assuming Arrokoth's geometric albedo of 0.21 applies to all cold classicals.

An important question relating to the possible formation of like-sized subunits in Arrokoth is whether different planetesimal masses (i.e., Arrokoth being far smaller than the 100 km characteristic size) imply correspondingly different mound-building subunits, or whether the subunits are more scale invariant, at least in a given region of the protosolar nebula. This can be tested in models, of course, but more definitive tests will come from the comparison of substantially larger KBOs and ancient planetesimals imaged by spacecraft or other means at high-enough resolution to discern the presence of mound structures like those seen on Arrokoth. Some useful work in this regard could be done by comparing images of cometary nuclei already examined by spacecraft, but these bodies are all smaller than Arrokoth. New missions to larger KBOs and Centaurs, as well as captured irregular satellites of the giant planets, would be highly useful in this regard, but such missions are necessarily far off in time as none are presently in construction or flight (although the Clipper and JUICE spacecraft may produce some new data on the Jovian irregular satellites). More immediately, we look forward to the imaging of the larger

NASA Lucy mission targets Eurybates (64 km diameter), Leucus (34 km diameter), Orus (51 km diameter), and Patroclus-Menoetius (113 and 104 km diameters, respectively).

# 10. Summary and Next Steps

Arrokoth's large mound structures dominate the appearance of its larger lobe Wenu, with tentative crypto-mounds having been now also identified on the limb of its smaller lobe, Weeyo. We believe that the self-similarity of the mounds argues for a common origin. We explore an exogenic origin of the mounds as accretionary precursors of Arrokoth that were assembled in the KBO's natal environment to create the body we see. How and why such a preferred size of ~5 km precursors should form Arrokoth, and how common this is in planetesimal formation, are open questions this study provokes for future research.

Because a return mission carrying new instruments to probe the interior and unseen hemispheres of Arrokoth is unlikely in any near-term future, we suggest other means to obtain relevant observations. These include:

➢ NASA Lucy mission observations of Trojan asteroids that may have formed similarly.
➢ ESA Comet Interceptor mission observations of an unevolved Oort Cloud comet.
➢ Future missions to other small KBOs, particularly CCKBOs.

Going forward, more detailed theoretical and experimental studies of KBO/planetesimal accretion are also of high interest, particularly at higher spatial resolution than has previously been available and via granular experiments in microgravity.

# Acknowledgements

This work was supported by the NASA New Horizons mission. We thank James T. Keane for an insightful review of this manuscript and Kirby Runyon for other helpful inputs. We thank Ms. Cynthia Conrad for editorial and submission assistance. We thank two anonymous referees for their helpful comments on this manuscript.